\documentclass[twocolumn,showkeys,showpacs,preprintnumbers,prd,superscriptaddress,nofootinbib]{revtex4}
\usepackage{graphicx}
\usepackage{epsf}
\usepackage{bm}
\usepackage{amsmath}
\usepackage{amsfonts}
\usepackage{amssymb}
\usepackage{epstopdf}
\usepackage{natbib}
\usepackage{hyperref}
\usepackage{color}
\usepackage{verbatim}
\usepackage{multirow}
\usepackage{bm}
\usepackage{hyperref}
\usepackage{listings}
\usepackage{soul}

\begin{document}

\title{A Hubble Constant Estimate from Galaxy Cluster and type Ia SNe Observations}

\author{L. R. Cola\c{c}o}
\email{colacolrc@gmail.com}
\affiliation{Instituto Federal da Paraíba (IFPB) - Campus Catolé do Rocha, 58884-000, Catolé do Rocha - PB, Brazil.}

\author{Marcelo Ferreira}
\email{fsm.fisica@gmail.com}
\affiliation{Universidade Federal do Rio Grande do Norte, Departamento de F\'{i}sica Te\'{o}rica e Experimental, 59300-000, Natal - RN, Brazil.}

\author{R. F. L. Holanda}
\email{holandarfl@gmail.com}
\affiliation{Universidade Federal do Rio Grande do Norte, Departamento de F\'{i}sica Te\'{o}rica e Experimental, 59300-000, Natal - RN, Brazil.}

\author{J. E. Gonzalez} \email{javiergonzalezs@academico.ufs.br}
\affiliation{Departamento de F\'{i}sica, Universidade Federal de Sergipe, São Cristóvão, SE 49100-000, Brazil}

\author{Rafael C. Nunes}
\email{costa.nunes@ufrgs.br}
\affiliation{Instituto de F\'{i}sica, Universidade Federal do Rio Grande do Sul, 91501-970 Porto Alegre RS, Brazil}
\affiliation{Divis\~ao de Astrof\'isica, Instituto Nacional de Pesquisas Espaciais, Avenida dos Astronautas 1758, S\~ao Jos\'e dos Campos, 12227-010, SP, Brazil}

\begin{abstract}
\noindent

In this work, we constrain the Hubble constant parameter, $H_0$, using a combination of the Pantheon sample and galaxy clusters (GC) measurements from minimal cosmological assumptions. Assuming the validity of the cosmic distance duality relation, an estimator is created for $H_0$ that only depends on simple geometrical distances, which is evaluated from Pantheon and a GC angular diameter distance sample afterward. The statistical and systematic errors in GC measurements are summed in quadrature in our analysis. We find $H_0 = 67.22 \pm 6.07$ ${\rm km \ s^{-1} Mpc^{-1}}$ in $1\sigma$ confidence level (C.L.). This measurement presents an error of around 9\%, showing that future and better GC measurements can shed light on the current Hubble tension.

\end{abstract}

\keywords{}


\maketitle

\section{Introduction}

In modern cosmology, the Hubble constant, $H_0$, plays a central role in our understanding of the universe's expansion and age. It is a fundamental parameter that characterizes the rate at which the universe is expanding and is essential in theoretical modeling for understanding the age and dynamics of the universe. In recent years, there has been significant interest and debate surrounding the precise value of the Hubble constant. Observations using different methods have yielded slightly different results, leading to tensions in determining $H_0$ \cite{riess2023local,freedman2023progress}.

The most statistically significant disagreement is between the Planck-CMB estimate \cite{Planck:2018vyg}, assuming the standard $\Lambda$CDM model, and the direct local distance ladder measurements conducted by the SH0ES team \cite{Riess:2021jrx,Riess_2022,murakami2023leveraging}, reaching a significance of more than 5$\sigma$. Additionally, many other late-time measurements are in agreement with a higher value for the Hubble constant (see the discussion in \cite{DiValentino:2021izs}) and in tension with the Planck-CMB estimate, for example, the Megamaser Cosmology Project \cite{Pesce:2020xfe} that gives $H_0 = 73.9 \pm 3.0$ km/s/Mpc, or using the Surface Brightness Fluctuations \cite{Blakeslee:2021rqi} that find $H_0 = 73.3 \pm 2.4$ km/s/Mpc.  On the other hand, the lower value of $H_0$ inferred from the Planck-CMB data is instead in very good agreement with Baryon Acoustic Oscillations (BAO) $+$ Big Bang Nucleosynthesis (BBN) constraints \cite{Sch_neberg_2019,Schoneberg:2019wmt,Cuceu:2019for}, and other CMB experiments like ACTPolDR4 \cite{ACT:2020frw},  ACTPolDR6 \cite{qu2023atacama} and SPT-3G \cite{Dutcher:2021vtw,SPT-3G:2022hvq}. The lower value of $H_0$ is also predicted by galaxy clustering analyses \cite{D_Amico_2021,Ivanov:2019pdj}. Motivated by such discrepancies, unlikely to disappear completely by introducing multiple systematic errors, it has been widely discussed in the literature whether new physics beyond the standard cosmological model can solve the $H_0$ tension \cite{DiValentino:2021izs,Abdalla:2022yfr,Perivolaropoulos_2022}.

Additionally, measuring $H_0$ independently of CMB data and local distance ladder method is important. In that regard, geometric distance measurements using galaxy clusters (GC) have a long history in astrophysics (see \cite{2011ARAA..49..409A,2023hxga.book..123C} for reviews). GCs are the most massive and large gravitational structures in the universe in hydrostatic equilibrium (or close to that). These vast assemblies of galaxies, dark matter, and gas bound together by gravity play a crucial role in our understanding of cosmology and the large-scale structure of the universe. Probes as the gas mass fraction \cite{Mantz_2021} imply a Hubble constant of $H_0 = 72.2 \pm 6.7$ km/s/Mpc in $1\sigma$ C.L. (see ref. \cite{2020JCAP...09..053H}), as well as allow to constrain dark energy models \cite{2012JCAP...02..035H,2014ApJ...781L..38L,2015MNRAS.447..479W,2019JCAP...11..032H,2020JCAP...09..053H,2022EPJWC.25700046W,Zhang_2023}. In the last years, GCs data have also been used in several other cosmological contexts, such as: for tests of the cosmic distance duality relation \cite{2010ApJ...722L.233H,2012JCAP...06..022H,2020JCAP...12..019H,2019APh...108...57Y}, for tests of fundamental physics \cite{2016JCAP...05..047H,2021ApJ...922...19L,2019JCAP...03..014C}, etc.

Very recently, the authors of \cite{PhysRevD.107.023520} introduced a way of measuring $H_0$ from a combination of independent geometrical data sets and without calibration or choice of a cosmological model. Such a method was built on the cosmic distance duality relation assumption, allowing measuring $H_0$ from first principles. Briefly, they obtained that $H_0 \propto H_0 D_L (z)/ D_A (z)$. Then, by using measurements of the apparent magnitude of SNe Ia as sources of $D_L(z)$, the line of sight and transverse BAO data as sources of $H(z)D_A(z)$, combined with cosmic chronometers as a source of $H(z)$, a value of $H_0 = 69.5 \pm 1.7$ km/s/Mpc ($1\sigma$ C.L.) was found, showing that the Hubble constant can be measured at the percent level with minimal cosmological assumptions (see details in section \ref{Methodology}). Still in a manner independent of the cosmological model, the authors \cite{Liao:2020zko} also determined the Hubble constant precisely through Gaussian Process regression using strong gravitational lensing systems with type Ia supernovae. An improved cosmological model-independent method of determining the value of the Hubble constant was also recently proposed by \cite{Liu:2023ifz}. The determination of $H_0$ by methods with minimal cosmological assumptions has been recently developed in the literature by several authors \cite{Liao_2019,D_Agostino_2023,Bonilla_2021,Dinda_2023,Zhang_2020,li2023modelindependent,Avila_2023}.

In this work, we aim to constrain the $H_0$ parameter by using a combination of independent data sets: luminosity distance from the Pantheon SNe Ia sample and galaxy cluster angular diameter distance obtained from their Sunyaev-Zel'dovich effect (SZE) and X-ray observations. As it is largely known, the combination of X-ray and SZE observations of galaxy clusters provides a direct measurement of the cosmic distance scale, independent of any local calibrator. The methodology used here follows that previously developed in \cite{PhysRevD.107.023520} (see section II), which is fully independent of any cosmological model. The advantage of this method to obtain $H_0$ is that there is no need for local calibration nor the choice of a cosmological model to infer $H_0$, but only assuming the validity of the cosmic distance duality relation (CDDR) \cite{Bassett:2003vu}. To date, there is no significant statistical evidence of the CDDR violation \cite{2010ApJ...722L.233H,2013MNRAS.436.1017L,2015JCAP...10..061S,2018MNRAS.480.3117L}. Thus, the method presents a robust and innovative way to infer constraints on the $H_0$ parameter. By using galaxy cluster diameter angular distances plus type Ia supernovae from Pantheon sample, we find $H_0 = 67.22 \pm 6.07$ ${\rm km \ s^{-1} Mpc^{-1}}$ with 1$\sigma$ C.L., representing a new measurement of $H_0$ with 9\% uncertainty (the statistical and systematic errors in GC measurements are summed in quadrature in our analysis). As one may see, unlike the $H_0$ estimate from \cite{PhysRevD.107.023520}, our result only depends on two types of astronomical data: luminosity and angular diameter distances. In the next section \ref{Methodology} we present the methodology and data sets to be used to infer $H_0$. In section \ref{results} we discuss our main results. Finally, in section \ref{final} we close the work with our final remarks and perspectives.

\section{Methodology and Data}
\label{Methodology}

In this section, we will describe the main steps of our methodology, which will be applied later to determine a new measurement of $H_0$. As well known, if the photon number traveling along null geodesics in a Riemannian space-time between the observer and the source is conserved, then the angular diameter distance, $D_{A}(z)$, and the luminosity distance, $D_{L}(z)$, satisfy the following expression: 

\begin{equation}
\label{CDDR}
D_L(z) = (1+z)^2 D_A(z).
\end{equation}
Such expression is largely known as CDDR\footnote{This quantity is also known as the Etherington-Hubble relation, and relates the mutual scaling of cosmic distances in any metric theory of gravity where photons are assumed massless and propagate on null geodesics \cite{Renzi:2021xii}.}, valid in any metric theory of gravity. Thus, such generality places this relation as being of fundamental importance in observational cosmology, and any deviation from it might indicate the possibility of new physics or the presence of systematic errors in observations \cite{CDDR,Bassett:2003vu,Ellis2007}. This relation has been tested with several astronomical data and to date there is no significant statistical evidence for a  CDDR violation \cite{2012A&A...538A.131H,2022ApJ...939..115X,2019ApJ...885...70L,2021EPJC...81..903L}. 

Following \cite{PhysRevD.107.023520}, one can rewrite eq. (\ref{CDDR}) to obtain:

\begin{equation}
\label{H0}
H_0 = \frac{1}{(1+z)^2} \frac{[H_0 D_L (z)]^{\rm SN}}{[D_A (z)]^{\rm GC}},
\end{equation}
where $[H_0 D_L (z)]$ is the so-called unanchored luminosity distance. It is clear that to obtain $H_0$ one has to measure the unanchored luminosity distance [$H_0 D_L (z)$]$^{\rm SN}$ and the angular diameter distance $[D_A (z)]^{GC}$ at the same redshift $z$. Thus, we shall take advantage of supernova type Ia as standard candles in order to obtain [$H_0 D_L (z)$]$^{\rm SN}$ and the Sunyaev-Zel'dovich effect with X-ray surface brightness of galaxy clusters to obtain $[D_A (z)]^{GC}$. {As one may see, this relation is fully independent of a specific cosmological model.} We shall present more details in what follows.

\subsection{Angular diameter distance by Galaxy Cluster}

The Sunyaev-Zel'dovich Effect (SZE) together with X-ray surface brightness can be used to obtain the angular diameter distance to galaxy clusters \cite{1979A&A....75..322C}. {The measured temperature decrement $\Delta T_{SZE}$ of the CMB due to the SZE can be quantified as (for details see Refs.  \cite{1979A&A....75..322C,2002ARA&A..40..643C,2004Ap&SS.290..105B,1999PhR...310...97B}):}

\begin{equation}
    \Delta T_{SZ} (\theta) = f(\nu,T_e) \frac{k_B T_e T_0}{m_e c^2}\sigma_{T} \int n_e dl,
\end{equation}
where $n_e$ is the electronic density of the intracluster medium, $c$ is the speed of light in vacuum $m_e$ is the electron mass and $f(\nu, T_e)$ takes into account the SZE frequency shift and relativistic corrections \cite{Itoh_1998,Nozawa_1998,2005AA434811C,2020MNRAS.493.3274L}, $k_B$ is the Boltzmann constant, $T_e$ is the electronic temperature and  $T_0 = 2.7255$ K is the present-day temperature of the CMB \cite{2009ApJ...707..916F}. Also, it is assumed that the temperature of the cluster is constant and no variations in the temperature cause higher order relativistic corrections as described in \cite{2013MNRAS.430.3054C}. The integration is along the line-of-sight $dl=D_A d\theta$, where $\theta$ is the line-of-sight angular size.

On the other hand, the X-ray emission is due to thermal Bremsstrahlung, and the surface brightness is given by \cite{1986RvMP...58....1S}:

\begin{equation}
    S_X(\theta) = \frac{1}{4\pi} \frac{D_{A}^{2}}{D_{L}^{2}}\int \frac{dL_X}{dV}dl,
\end{equation}
where the emissivity in the frequency band $[\nu_1, \nu_2]$ is usually written as

\begin{equation}
    \frac{dL_X}{dV} =  \left(   \frac{2\pi k_B T_e}{3m_e} \right)^{1/2}  \frac{2^4 e^6}{3m_e \hbar c^3} n_e \left( \sum_i Z_{i}^{2}n_i g_{B_i}    \right),
\end{equation}
$Z_i$ and $n_i$ are, respectively, the atomic numbers and the distribution of elements,  $g_B$ is the Gaunt factor which takes into consideration the corrections due to quantum and relativistic effects of Bremsstrahlung emission, $\hbar$ is the Planck constant (divided by 2$\pi$) and $e$ is the electron charge.  Considering the intracluster medium basically constituted of hydrogen and helium, one may write \cite{1986RvMP...58....1S}:

\begin{equation}
    \frac{dL_X}{dV}  \approx 3.10^{-27}T_e^{1/2} n_e^2 \: \: \: \text{ergs} \: \text{cm}^{-3}\: \text{s}^{-1},
\end{equation}
considering $T_e$ and $n_e$  the numerical values of temperature and density measured in K and cm$^{-3}$, respectively. From now on we will use $\frac{dL_X}{dV} = \Lambda_e n_e^{2}$ \cite{1986RvMP...58....1S}.

To illustrate the SZE/X-ray method, we will employ the spherical isothermal $\beta$-model, which describes electronic density. This model, though simplistic, is widely referenced in the literature due to its analytical solutions for integrals. While it facilitates a quantitative demonstration of the methodology, it is important to note that it serves primarily for illustrative purposes. Then, under this assumption the electronic density is given by \cite{1978AA....70..677C}:

\begin{equation}
\label{modelbeta}
    n_e(\theta) =  n_0    \left[  1 +  \frac{\theta^2}{\theta_c^2} \right]^{-3 \beta/2}.
\end{equation}
Then, by integrating the equation (3), one may obtain \cite{1979A&A....75..322C,2002ARA&A..40..643C,2004Ap&SS.290..105B,1999PhR...310...97B}:

\begin{equation}
 \Delta T_{SZ} = \Delta T_0  \left(   1+ \frac{\theta^{2}}{\theta_c^2}  \right)^{\frac{1-3\beta}{2}},
\end{equation}
where $\theta_c$ is the corresponding angular core radius  and  $\Delta T_0$ is the central temperature decrement. More precisely:

\begin{equation}
\label{ESZ}
\Delta T_0 \equiv T_0 f(\nu, T_e) \frac{\sigma_T k_B T_e}{m_e c^2} n_{0} \sqrt{\pi} \theta_c D_A g(\beta/2),
\end{equation}
where $g(\alpha) \equiv \frac{ \Gamma \left[ 3\alpha - 1/2 \right]}{\Gamma [3\alpha ]}$, and $\Gamma(\alpha)$ is the gamma function.

On the other hand, for X-ray surface brightness, we may obtain:

\begin{equation}
    S_X = S_{X_0} \left(  1+ \frac{\theta^2}{\theta_{c}^{2}}   \right)^{1/2 - 3\beta},
\end{equation}
whereas the central surface brightness ($S_{X_0}$) is:

\begin{equation}
\label{xray}
S_{X_0} \equiv \frac{D_{A}^{2}\Lambda_e}{4D_{L}^{2}\sqrt{\pi}} n_{0}^{2} \theta_c D_A g(\beta).
\end{equation}

So, by solving equations \ref{ESZ} and \ref{xray}  by eliminating $n_0$ and taking into consideration the validity of the CDDR, it is possible to obtain:

\begin{eqnarray}
[D_A(z)]^{\rm GC} = \left[  \frac{\Delta T_{0}^{2}}{S_{X_0}} \left( \frac{m_e c^2}{k_B T_e}\right)^2 \frac{g(\beta)}{g(\beta/2)^2}   \right]   \nonumber \\ 
\times \frac{\Lambda_e}{4 \pi^{3/2} f(\nu, T_e)^2 T_{0}^{2} \sigma_{T}^{2} \theta_c (1+z_c)^4 },
\end{eqnarray}
from which we can estimate galaxy cluster angular diameter distance measurements (see, for instance, the Ref. \cite{2002ApJ...581...53R}).

However, the usually assumed spherical geometry for galaxy clusters has been severely questioned after analyses based on data from the XMM-Newton and Chandra satellites had suggested that clusters exhibit preferably an elliptical surface brightness. Additionally, galaxy cluster angular diameter distance data obtained by the spherical assumption has also been rejected in several tests of the CDDR \cite{2010ApJ...722L.233H,2011A&A...528L..14H,2012ApJ...745...98M,2015APh....62..134H}. Moreover, in  Ref. \cite{2009MNRAS.396..709W}, by using simulations, it  was predicted  that dark matter halos show axis ratios typically of the order of
$\approx 0.8$.  It is worth noting that the first determination of the
intrinsic three-dimensional (3D) shapes of galaxy clusters was presented in the Ref. \cite{2010ApJ...713..491M}  by combining X-ray, weak-lensing, and strong-lensing observations. The galaxy cluster MACS J1423.8+2404 was analyzed  and the authors found a tri-axial galaxy cluster geometry with Dark matter halo axial ratios of $1.53
\pm 0.15$ and $1.44 \pm 0.07$ on the plane of the sky and along the line of sight, respectively.

In our analyses, we use the galaxy cluster sample of $D_A$ compiled by \cite{DeFilippis_2005} (see our Fig.1 (left)). This sample is composed of 25 galaxy clusters ($[D_A]^{GC}(z)$) in the redshift range $0.023 < z_{GC} < 0.784$, { where an isothermal elliptical $\beta$-model  was considered to describe the main cluster properties and obtain the angular diameter distances.  Actually,  the angular diameter distance to clusters presented in Ref. \cite{DeFilippis_2005} was obtained in three ways by using their Redshifts: considering the cosmic concordance model (second column of their Table I), by using the SZE/X-ray technique with spherical assumption (third column of their Table I) and by using the SZE/X-ray technique with elliptical assumption (fourth column of their Table I). For this case, the authors assumed the clusters are oblate spheroids. We have used the fourth.} \footnote{ In appendix A from \cite{DeFilippis_2005} it is possible to find a detailed discussion on the elliptical $\beta$-model. }.

As a matter of clarification, the authors of  \cite{DeFilippis_2005} showed, for this sample, that the choice of circular rather than the elliptical model does not affect the resulting central surface brightness or Sunyaev-Zel'dovich decrement and the slope $\beta$ differs moderately between these models. However, different values for the projected core radius can be obtained. More specifically,  they verified that the obtained projected core radius by using the spherical assumption is lower than that from the elliptical one (see Fig.(1) of \cite{DeFilippis_2005}). Therefore, the assumed cluster shape can affect considerably the SZE/X-ray distances, and, consequently, the $H_{0}$ estimates and other astrophysical quantities. In a first approximation, it is possible to correct the projected core radius from the spherical assumption by using  $\theta^{\rm spher.}_{\rm corr} = \frac{2e_{\rm proj}}{1+e_{\rm proj}}\theta_{\rm circ}$ \cite{DeFilippis_2005}, where  $\theta^{\rm spher.}_{\rm corr}$ and $\theta_{\rm circ}$ are the core radius corrected and that one from the isothermal spherical $\beta$ model, respectively. The factor $e_{\rm proj}$ is the axial ratio of the major to the minor axes of the projected isophotes.

\begin{figure*}
\label{DAsamples}
    \includegraphics[scale=0.47]{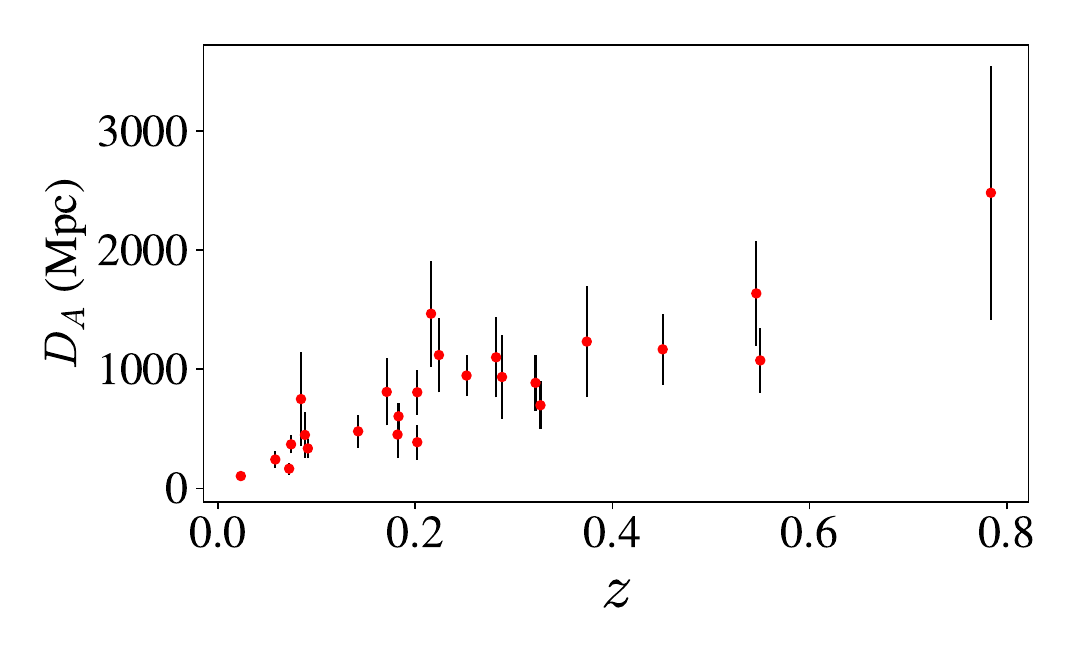}
    \includegraphics[scale=0.47]{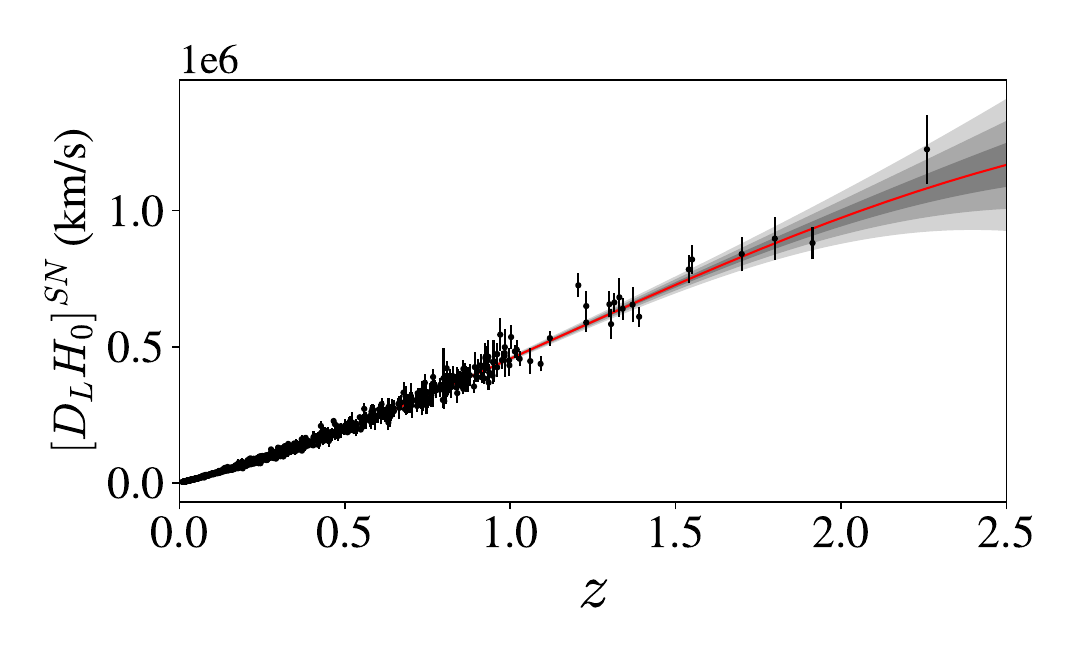}
    \caption{Left Panel: Angular diameter distances to galaxy clusters.
Right Panel: Reconstruction of $[H_0 D_L]^{\rm SN}(z)$ from the Pantheon sample. The solid red line and the shadowed regions represent the reconstruction central value, and the 1$\sigma$, 2$\sigma$, and 3$\sigma$ C.L. obtained by the GP method.}
\end{figure*}

It is important to stress that for the galaxy cluster observations, the common statistical contributions are: SZE point sources $\pm 8\%$, X-ray background $\pm 2\%$, Galactic $N_H$ $\leq \pm 1\%$, $\pm 15\%$ for cluster asphericity, $\pm 8\%$ kinetic SZE and for CMB anisotropy $\leq \pm 2\%$.  Estimates for systematic effects are: SZE calibration $\pm 8\%$, X-ray flux calibration $\pm 5\%$, radio halos $\pm 3\%$ and X-ray temperature calibration $\pm 7.5\%$ (see, for instance, Table 3 \cite{Bonamente_2006}). In the present analysis, we have combined the statistical and systematic errors in quadrature for the angular diameter distance from galaxy clusters. Therefore, the left panel of Figure 1 shows the $[D_{A}]^{GC}(z)$ sample used in this paper. Through the years, this data set has been used in several {cosmological contexts}, such as: in cosmological parameter estimates \cite{2012JCAP...02..035H,2013JCAP...06..033H,2014JCAP...02..021L,2014ApJ...781L..38L}, for tests of the cosmic distance duality relation \cite{2010ApJ...722L.233H,2013MNRAS.436.1017L,2015JCAP...10..061S,2018MNRAS.480.3117L}, for tests of fundamental physics \cite{2016JCAP...05..047H,2021ApJ...922...19L,2019JCAP...03..014C}, etc. 

\subsection{The unanchored luminosity distance}

To obtain the absolute distance ($D_L(z)$) from SNe Ia, one needs the absolute magnitude, $M_B$. However, observations of SNe Ia can also provide the so-called unanchored luminosity distance $[H_0 D_L (z)]$, which is the quantity we shall use for our purposes (see details in ref. \cite{PhysRevD.107.023520}). Thus, the unanchored luminosity distances are derived from the apparent magnitude of SNe Ia by the relation:

\begin{equation}
\label{m_B}
m_B = 5 \log_{10} [H_0 D_L (z)] - 5a_B,
\end{equation}
where $m_B$ is the apparent magnitude measurements, and we assume the fixed value of $a_B = 0.71273 \pm 0.00176$ as inferred by \cite{Riess:2016jrr}. Such value is independent of any absolute scale, luminosity, or distance, and is determined from a Hubble diagram of SNe Ia with a light-curve fitter.

For our purpose, we shall use the Pantheon type Ia supernovae sample in order to obtain the unanchored luminosity distances, $[H_0 D_L]^{\rm SN} (z)$, at the redshift of the GC sample. The Pantheon compilation consists of 1048 measurements of SNe Ia apparent magnitudes, $m_b$, in the redshift range $0.01 < z < 2.3$ \cite{Scolnic:2017caz}. Then, we transform the Pantheon sample of apparent magnitudes into a sample of unanchored luminosity distances by considering the relation (\ref{m_B}) via \cite{PhysRevD.107.023520}:

\begin{equation}
\label{unanchored_luminosity}
[H_0 D_{L}]^{\rm SN}(z) = 10^{(m_b + 5a_B)/5}\equiv 10^{m_b'/5}.
\end{equation}

To estimate the $[H_0 D_{L}]^{\rm SN}$ uncertainties including their correlations, we consider the covariance matrix of the apparent magnitudes (statistics + systematics) and the $a_B$ error. The covariance matrix of $m'_b$ is obtained by \footnote{Quantities in bold represent vectors or matrices.}:
\begin{equation}
    \textbf{Cov} (\bm m'_b)=\textbf{Cov}(\bm m_b)+(5\sigma_{a{_B}})^2\bm I,
\end{equation}
\noindent where $\bm I$ is the unity matrix. The  covariance of the luminosity distances is calculated using the matrix transformation relation as follows 
\begin{equation*}
\textbf{Cov}([\bm {H_0 D_{L}}]^{SN}) =
\end{equation*}
\begin{equation}
\label{error_unanchored_luminosity}
\left(\frac{\partial  [\bm H_0 \bm D_{L}]^{SN} }{\partial \bm m'_b}\right)\textbf{Cov}(\bm m'_b) \left(\frac{\partial  [\bm H_0 \bm D_{L}]^{SN} }{\partial \bm m'_b}\right)^T,
\end{equation}
where $\frac{\partial  [\bm H_0 \bm D_{L}]^{SN} }{\partial \bm m'_b}$ represents the partial derivative matrix of the unanchored luminosity distance vector $[\bm H_0 \bm D_{L}]^{SN} $ concerning the vector $\bm m'_b$. It is worth commenting that the redshift of a galaxy cluster represents the average redshift of its constituent galaxies, naturally subject to some degree of uncertainty. However, in a massive galaxy cluster, the typical velocity dispersion among galaxies is approximately 1000 km/s. Consequently, the redshift error stemming from the uncertainty in this mean velocity is given by $\Delta z \approx \Delta v/c = \sigma_v/c\sqrt{n}=0.0033/\sqrt{n}$, where $n$ denotes the number of galaxies within the cluster with measured redshifts \cite{2006eac..book.....S}. Therefore, this error in $z$ within CG samples can be considered negligible. Also, as well known, errors in redshift measurements for SN are negligible. As a result, no error bars are associated with the variable $z$. Thus, the variable $z$ is continuously varied in all Gaussian Process conducted across all sample data.

In addition, we use the Gaussian Process (GP) method trained on the SNe Ia sample, to reconstruct the $[H_0 D_L (z)]^{\rm SN}$ function (refer to Fig. \ref{DAsamples}, right panel). The trained network shall be able to forecast unanchored luminosity distances at redshifts that match $[D_A (z)]^{GC}$ from galaxy clusters. 

The GP reconstruction is performed by choosing a prior mean function and a covariance function which quantifies the correlation between the values of the dependent variable of the reconstruction and is characterized by a set of hyperparameters (see ref. \cite{Seikel:2013fda} for more details about GP). In our reconstructions of $[H_0 D_L(z)]^{\rm SN}$, we choose zero as the prior mean function to avoid biased results and a Gaussian kernel as the covariance between two data points separated by a $z - z'$ redshift distance given by:

\begin{equation}
\label{gaussian_kernel}
k(z,z')=\sigma^2 \exp\left(-\frac{(z-z')^2}{2l^2}\right),
\end{equation}
where $\sigma$ and $l$ are the hyperparameters related to the variation of the estimated function and its smoothing scale, respectively. To estimate the hyperparameter values, we maximize the logarithm of the marginal likelihood:

\begin{equation*}
\ln {\mathcal{L}}=-\frac{1}{2}[\bm {H_0 D_{L}}]^{SN} [\bm K(\bm z,\bm z)+\bm C]^{-1}([\bm {H_0 D_{L}}]^{SN})^T
\end{equation*}

\begin{equation}
\label{log_likelihood}
  -\frac{1}{2}\ln |\bm K(\bm z,\bm z)+\bm C|-\frac{n_d}{2}\ln 2\pi,
\end{equation}

\noindent where $\bm{ z}$ is the vector of redshift measurements of the Pantheon data, $\bm K(\bm z,\bm z)$ is the covariance matrix used to describe the data as a GP and its elements are calculated with Eq. (\ref{gaussian_kernel}) ($[\bm K(\bm z,\bm z)]_{i,j}=k(z_i,z_j)$), $\bm C$ is the covariance matrix of the data ($[H_0 D_L (z)]^{SN}$ error matrix) obtained using the Eq. (\ref{error_unanchored_luminosity}) and $n_d$ is the number of data points.

We use the code GaPP code\footnote{https://github.com/carlosandrepaes/GaPP} to perform GP reconstruction of the $[H_0 D_L (z)]^{SN}$ function. In what follows, we discuss our main results.

\section{Main Results}
\label{results}

We use Markov Chain Monte Carlo (MCMC) methods to estimate the posterior probability distribution functions (pdf) of free parameters supported by the \texttt{emcee} MCMC sampler \cite{2013PASP125306F}. To perform the plots, we used the \texttt{GetDist} Python package \cite{Lewis:2019xzd}. The likelihood is given by:

\begin{equation}
    \mathcal{L}(\bm H_{0}^{\mathrm{Data}}|H_0) =  \frac{\exp{{ (-\chi^2/2)}}}{(2\pi)^{n_d/2}|\bm C_{H_0}|^{1/2}} ,
\end{equation}
with

\begin{equation}
    \chi^2 = (\bm H_0 - \bm H_{0}^{\mathrm{Data}})\bm C_{H_0}^{-1}(\bm H_0 - \bm H_{0}^{\mathrm{Data}})^T,
\end{equation}
where $\bm H_{0}^{\mathrm{Data}}$ is the data vector given by eq.(\ref{H0}), $\bm H_0$ is the free parameter,  $\bm C_{H_{0}}^{-1}$ is the inverse of the covariance matrix, and $n_d$ is the number of Hubble constant data. The pdf is proportional to the product between likelihood and prior ($P(H_0)$), that is,

\begin{equation}
    P(H_0|\bm  H_{0}^{\mathrm{Data}})  \propto  \mathcal{L} (\bm H_{0}^{\mathrm{Data}}|H_0) \times P(H_0).
\end{equation}

\begin{figure}
\label{RESULTS}
    \includegraphics[scale=0.75]{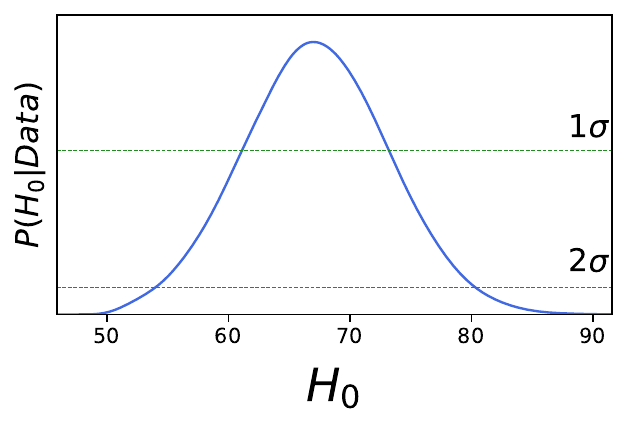}
    \caption{Posterior probability distribution functions of $H_0$. The green horizontal dashed lines correspond to $1\sigma$ and $2\sigma$ C.L., respectively.}
\end{figure}

\noindent In our analyses, we assume flat prior: $H_0 = [50, 100]$ km/s/Mpc.

We combined the 25 points of $[D_A(z)]^{\rm GC}$ with SNe Ia in order to obtain $H_0$. After summing the statistical and systematic errors in GC measurements in quadrature. By assuming gaussianity in the systematic and statistical errors, the data remains gaussian with resulting variance equal to the sum of the statistical and systematic variances. By employing this method, we can comprehensively address the cumulative impact of uncertainties, thereby yielding a more precise assessment of the overall uncertainty inherent in a given measurement. In our analysis, we obtain $H_0 = 67.2 \pm 6.1$ ${\rm km \ s^{-1} Mpc^{-1}}$ at the $1\sigma$ C.L. with an error around 9\% (see figure \ref{RESULTS}). Due to errors presented in galaxy cluster observations, our method provided consistent and compatible results of the Hubble parameter at low and intermediate redshifts. Moreover, this constraint is in full agreement with the one from the Planck satellite observations, but it is also compatible with the measure provided by the SH0ES team. However, it is very important to stress that unlike the $H_0$ estimate from \cite{PhysRevD.107.023520}, our result only depends on two types of observations, moreover, the $[D_A]^{\rm GC}$ quantity used here is completely independent of local calibrators.

We also estimated the value of $H_0$ using data from De Filippis et al. \cite{DeFilippis_2005}, with the assumption of the spherical model to describe the galaxy clusters (third column of table III in their paper). We obtain $H_0 = 62.53 \pm 2.3$ ${\rm km \ s^{-1} Mpc^{-1}}$ at $1\sigma$ C.L., which is in stark disagreement with values obtained from CMB and SH0ES measurements.
Previous estimates of $H_0$ using this model have consistently yielded low values, as reported in some cosmological references \cite{1997ApJ...480..449H,2002ApJ...581...53R}. Clearly, this simplification in the morphology of the GC leads to biased results on $H_0$. Therefore, our results also suggest that the elliptical model is more realistic in describing galaxy clusters, as previously anticipated and well described in the literature.

\section{Final Remarks}
\label{final}

In this paper, we discussed a determination of the Hubble constant $H_0$, without using any cosmological model, based only on the SZE/X-ray distance technique for measuring galaxy cluster angular diameter distances and type Ia supernovae observations. We followed the method proposed in \cite{PhysRevD.107.023520} (see \ref{Methodology}), but only two types of astronomical measurements were considered here. In galaxy cluster observations (25 data), the electronic density of the intracluster medium was described by an isothermal elliptical $\beta$-model \cite{DeFilippis_2005}. For SNe Ia, we used the Pantheon sample. It was obtained $H_0 = 67.2 \pm 6.1$ ${\rm km \ s^{-1} Mpc^{-1}}$ at $1\sigma$ C.L. It is important to comment that the statistical and systematic errors in galaxy cluster measurements are summed in quadrature in our analysis. As one may see, the constraints on the Hubble constant derived here are consistent with the latest and conflicting measures provided by the SH0ES team and the Planck satellite. This occurs due to errors present in galaxy cluster observations.

Finally, it is worth noting that the galaxy cluster angular diameter determined from their SZE + X-ray observations provides an alternative method for determining direct cosmological distance. Our results further underscore the importance of actively pursuing observational efforts to obtain new galaxy cluster angular diameter distance samples through SZE and X-ray observations. With the advent of larger and more precise datasets characterized by reduced statistical and systematic uncertainties, the methodology employed in this study holds the potential to refine constraints on the Hubble parameter. Simultaneously, relativistic correction coming from our own motion \cite{2005A&A...434..811C}, may become feasible taken into account for modeling SZ signal \cite{2012MNRAS.426..510C}. We look forward to exploring these avenues in future communications.

\begin{acknowledgments}
\noindent The authors express their gratitude to the referee for providing valuable comments and suggestions aimed at enhancing the overall quality of the work. RFLH thanks to CNPQ support under the project No. 309132/2020-7. RCN thanks the financial support from the Conselho Nacional de Desenvolvimento Científico e Tecnologico (CNPq, National Council for Scientific and Technological Development) under the project No. 304306/2022-3, and the Fundação de Amparo à Pesquisa do Estado do RS (FAPERGS, Research Support Foundation of the State of RS) for partial financial support under the project No. 23/2551-0000848-3.
\end{acknowledgments}

\bibliography{PRD}

\end{document}